\documentclass[11pt,a4paper]{article}
\usepackage{graphicx}
\usepackage{amssymb}
\usepackage{amsmath}
\usepackage{cite}
\title{Average size of random polygons with fixed knot topology}
\author{$^1$Hiroshi Matsuda, $^1$Akihisa Yao, $^2$Hiroshi Tsukahara,\\
$^3$Tetsuo Deguchi, $^4$Ko Furuta and $^1$Takeo Inami
\vspace{3mm}\\
$^1$Department of Physics, Faculty of Science and Engineering,\\
Chuo University,
1-13-27 Kasuga, Bunkyo-ku, Tokyo 112-8551\\
\\
$^2$Geographic Infomation Systems Department,\\
Hitachi Software Engineering Co., Ltd.\\
4-12-7 Higashishinagawa, Shinagawa-ku, Tokyo 140-0002\\
\\
$^3$Department of Physics,\\
Faculty of Science, Ochanomizu University\\
2-1-1 Ohtsuka, Bunkyo-ku, Tokyo 112-8610\\
\\
$^4$Department of Physics, National Tsing Hua University\\
101 Section 2 Kuang Fu Road, Hsinchu, Taiwan 300, R.O.C.}
\date{22 March 2003}

\begin{document}
\maketitle

\begin{abstract}
We have evaluated by numerical simulation the average size $R_K$ of random polygons of
fixed knot topology $K = \emptyset, 3_1, 3_1\sharp4_1$, and we have confirmed the scaling law 
$R^2_K \sim N^{2\nu_K}$ for the number $N$ of polygonal nodes in a wide range; $N = 100$ -- $2200$.
The best fit gives $2\nu_{K} \simeq 1.11$ -- $1.16$ with good fitting curves in the whole range of $N$.
The estimate of $2 \nu_K$ is consistent with the exponent of self-avoiding polygons.
In a limited range of $N$ ($N \gtrsim 600$), 
however, we have another fit with $2 \nu_K \simeq 1.01$ -- $1.07$, 
which is close to the exponent of random polygons.
\end{abstract}

%
\section{Introduction}
Topology of polymers is an important issue in understanding the
physical properties of polymer materials, 
such as the viscoelasticity of polymer solutions~\cite{DOI-EDWARDS.1986.01}.
It is also relevant to fundamental problems in biology~\cite{DELBRUCK.1962.01};
the enzymes can cut and reconnect DNA strands,
which are crucial functions in the mechanism of the complex processes 
of reproduction, transcription and recombination of DNA strands~\cite{SUMNERS.1995.01}.
The topology of DNA rings can be used as a probe to detect the action
of the enzymes on the DNA strands by observing the change of their knot type.
The recent experimental observation~\cite{STASIAK-ET-AL.1996.01} 
reports that the average polymer size depends
on the topology of polymers.

One important quantity in the study of ring polymers is 
their average size with fixed knot type.
This quantity has been estimated in several numerical methods~\cite{
DES_CLOIZEAUX-METHA.1979.01,QUAKE.1994.01,
JANS_VAN_RENSBURG-WHITTINGTON.1991.01,ORANDINI-TESI-JANSE_VAN_RENSBURG-WHITTINGTON.1998.01,
DEUTSCH.1999.01,
SHIMAMURA-DEGUCHI.2001.02,SHIMAMURA-DEGUCHI.2001.03,SHIMAMURA-DEGUCHI.2002.01}.
The renormalization group (RG) argument leads to the power law scaling for the average size
of linear polymers  
as thier length increases~\cite{DE_GENNES.1972.01,WILSON.1974.01,
DES_CLOIZEAUX.1974.01},
\begin{equation}
R^2(N) = AN^{2\nu}\left( 1 + BN^{-\Delta} + \dots \right),
\label{eq:scaling-all}
\end{equation}
where $N$ is the number of segments,
 and $R^2(N)$ is the mean square size of polymers.
The same scaling law should hold for ring polymers without topological constraint 
with the same value of the exponent $\nu$.
It is commonly accepted that the size of ring polymers 
with fixed knot topology $K$ is given by the same scaling relation 
as (\ref{eq:scaling-all}),
\begin{equation}
R^2_{K}(N) = A_{K}N^{2\nu_{K}}\left( 1 + B_{K}N^{-\Delta_K} + \dots \right).
\label{eq:scaling-K}
\end{equation}
This conjecture is supported by numerical simulations~\cite{
JANS_VAN_RENSBURG-WHITTINGTON.1991.01,
ORANDINI-TESI-JANSE_VAN_RENSBURG-WHITTINGTON.1998.01,
DEUTSCH.1999.01,SHIMAMURA-DEGUCHI.2001.02,SHIMAMURA-DEGUCHI.2001.03,
SHIMAMURA-DEGUCHI.2002.01}.

The scaling law of the form (\ref{eq:scaling-all}) 
has already been discussed in connection with random walks (RW) 
and self-avoiding walks (SAW).
Closed paths of RW and SAW are respectively random polygons and self-avoiding polygons (SAP).
The exponents in these two cases are well established: 
$\nu_{RW} = {1 \over 2}$ and $\nu_{SAW} \simeq 0.588$~\cite{
LIPKIN.1981.01,PRENTIS.1981.01}.
The central question regarding the scaling (\ref{eq:scaling-K}) is: 
What is the value of $\nu_K$ ?
Is $\nu_K$ related to either $\nu_{RW}$ or $\nu_{SAW}$?
Note that the exponent $\nu_K$, or the coefficient $A_K$ or $B_K$,
should depend on the knot type $K$.

It has been found by numerical analysis that, in the case of SAP, 
the exponent $\nu_K$ does not depend on the knot type $K$, 
and is given by $\nu_{SAW}$~\cite{
JANS_VAN_RENSBURG-WHITTINGTON.1991.01,
ORANDINI-TESI-JANSE_VAN_RENSBURG-WHITTINGTON.1998.01}.
The question is then whether 
the effect of topology appears in $A_K$ or at 
the level of correction to the scaling (\ref{eq:scaling-K}),
i.e. in $B_K$~\cite{WHITTINGTON.1992.01}.
Simulations by van Rensburg et al.~\cite{
JANS_VAN_RENSBURG-WHITTINGTON.1991.01,
ORANDINI-TESI-JANSE_VAN_RENSBURG-WHITTINGTON.1998.01} 
support the conjecture that 
knot topology does not affect the coefficient $A_K$.

For the Gaussian model of polymers, 
des Cloizeaux made the conjecture that the constraint of topology should lead effectively to
the growth of their average size, so called the "topological excluded volume effect",
even though it has no excluded volume interactions~\cite{DES_CLOIZEAUX.1981.01}.
It is predicted that the exponent $\nu_\emptyset$ for polymers with trivial knot topology
$\emptyset$ obeys $\nu_\emptyset \le \nu_{SAW}$.
Deutsch has performed a simulation for phantom chains in the case of 
trivial knot $\emptyset$ and 
obtained the value of $\nu_\emptyset$ 
consistent with this inequality~\cite{DEUTSCH.1999.01}.
The conjecture is also supported by a phenomenological argument~\cite{GROSBERG.2000.01} 
for $\emptyset$ and other knots.

Recently, Shimamura and Deguchi have evaluated the exponent $\nu_K$ for the radius of gyration 
for knot types $\emptyset$, $3_1$ and $4_1$ 
in the case of the Gaussian random polygons~\cite{
SHIMAMURA-DEGUCHI.2001.03}.
Taking into account the correction term in the scaling (\ref{eq:scaling-K}), 
they have found that the fit gives $\nu_K \simeq \nu_{RW}$.

In this paper, we re-examine the radius of gyration of the closed phantom chain model.
For the topological effect on the size of random polygons,
the two different answers have been presented.
Furthermore, other possibilities are not excluded.
The topological effect should also be investigated in different models.
We employ the pivot algorithm which is a modification of 
the original one~\cite{DEUTSCH.1999.01} so that the possible bias is reduced.
The simulation procedure used in~\cite{
DEUTSCH.1999.01}, which is a kind of pivot algorithm~\cite{
MADRAS-SOKAL.1987.01,MADRAS-ORLITSKY-SHEPP.1990.01}
for continuum models, tends to pick up extended conformation 
more frequently.
It is conceivable that this bias leads to an artificial expansion of chains 
and to raise the exponent $\nu_K$ from $\nu_{RW}$.
We have evaluated the radii of gyration for the range of chains
$N = 100$ -- $2200$ for three knot types, 
trivial knot, trefoil knot and composite knot, denoted by 
$\emptyset$, $3_1$ and $3_1\sharp4_1$.
In the case of $\emptyset$, our improved algorithm has given 
a result consistent with that of~\cite{DEUTSCH.1999.01}.
We have further results in other cases.

We have found two good fits to our simulation data.
The best fit gives $2\nu_{K} \simeq 1.11$ -- $1.16$.
It is consistent with $\nu_{SAW}$,
which also agrees with the result of Deutsch:
$\nu_{\emptyset} \simeq 1.17$~\cite{DEUTSCH.1999.01}.
The second fit gives $2\nu_K \simeq 1.01$ -- $1.07$,
which is in accordance with $\nu_{RW}$.
The first fit gives good fitting curves for the whole range of $N$
investigated in simulation: $N = 100$ -- $2200$.

%
\section{Model}
We consider closed phantom chains in three dimensions, 
consisting of $N$ line segments of length $a$.
We call them $N$-node polygons.
We assume that the polygons have no excluded volume.
It means that the nodes and segments of the polygon are
purely geometrical points and line segments, respectively
~\cite{DEUTSCH.1999.01}.
This model may be regarded as a model for polymers at 
the $\theta$-point or a polymer in a melt.

A polygon ${\mathcal P}_N$ is defined by the set of position vectors
of its nodes, ${\mathcal P}_N = \left( R_1, R_2, \dots, R_N \right)$.
All cyclic permutations of one position vectors 
correspond to the same polygon.
The vectors satisfy the geometrical constraint 
$|R_{i+1}-R_{i}| = a$ for $1 \leq i \leq N$ ($R_{N + 1} = R_1$).
Each polygon is topologically equivalent to a knot $K$ in three dimensions.
The configuration space $\mathcal{C}$ of the model is divided into subspaces 
$\mathcal{C}_K$ in which all polygons have a fixed knot type $K$,
$\mathcal{C} = \sum_K \mathcal{C}_K$.

The radius of gyration of polygons is given by
\begin{equation}
R^2({\mathcal P}_N) = \frac{1}{2N^2} \sum_{i,j=1}^{N} \left( R_i - R_j \right)^2.
\end{equation}
This definition is different from that of the polygon size employed
in~\cite{DEUTSCH.1999.01}
but their asymptotic behaviors in the limit of large $N$ are the same.
We set the segment length $a$ to unity; 
macroscopic properties are independent of the microscopic parameter $a$.

We generate a large number $M$ of polygons with length $N$. 
They are used to evaluate the mean square size of polygons 
without topological constraint 
\begin{equation}
R^2(N) = \frac{1}{M} \sum_{i = 1}^{M} R^2({\mathcal P}_{N,i}).
\label{eq:all}
\end{equation}
and the same quantities for fixed knot topology, 
\begin{equation}
R^2_{K}(N) = \frac{1}{M_K} \sum_{i = 1}^{M} R^2({\mathcal P}_{N,i}) 
\chi({\mathcal P}_{N,i},K),
\label{eq:k}
\end{equation}
Here the indicator function $\chi({\mathcal P},K)$ takes the value $1$ 
if ${\mathcal P} \in \mathcal{C}_K$ and zero otherwise. 
The number of polygons $M_K$ with fixed topology $K$ is 
$M_K = \sum_i \chi({\mathcal P}_{N,i}, K)$.

We are concerned with the scaling relation
\begin{equation}
\frac{R^2_{K}}{R^2}=\frac{A_{K}}{A}
N^{2\Delta\nu_K}\left[1+\Delta B_K
N^{-1/2}+O(N^{-1})\right] ,
\label{eq:par}
\end{equation}
where $\Delta\nu_K = \nu_{K}-\nu$, $\Delta B_K = B_{K}-B$.
It follows from the scaling laws (\ref{eq:scaling-all}) and 
(\ref{eq:scaling-K}) with
$\Delta = \Delta_K = 1/2$~\cite{ORANDINI-TESI-JANSE_VAN_RENSBURG-WHITTINGTON.1998.01}.
We perform the fit simulation data using (\ref{eq:scaling-K})
%
\section{Simulation Procedure}
Our sampling of the polygons follows the dynamic Monte Carlo
method using the pivot algorithm
~\cite{MADRAS-SOKAL.1987.01,MADRAS-ORLITSKY-SHEPP.1990.01},
applied to the continuum model by Deutsch~\cite{DEUTSCH.1999.01}.
In the continuum model, a pivot move for a polygon is a rotation
of a chain of segments randomly chosen from the polygon 
around the axis passing the two endmost nodes of the chain
by a random amount of angle $\theta$.

In practice the method used in~\cite{DEUTSCH.1999.01} imposes 
two restrictions on pivot moves, 
1) the banning of self-intersections during the move and 
2) the range of angle $\theta$.
There is a possibility that they may cause a virtual expansion of polygons 
due to these restrictions.
The transformation process of the pivot move 
prohibits those moves in which self-intersections of 
polygonal segments occur.
The angle $\theta$ is selected from the range of 
$-90^{\circ} \le \theta \le 90^{\circ}$, 
excluding rotations with $\theta > 90^{\circ}$.
This tends to bring polygons to less folded conformations.
and hence the polygon size is increased.
Furthermore, the rotations are not completely randomly distributed 
over the range of $\theta$, since the rotations are limited to 
those which make no change of the knot topology.

In this paper, we modify the algorithm in two respects: 
We allow the selected chain of segments to rotate by an angle $\theta$
between $0$ and $360$ degrees.
We do not check the self-intersections during the process of rotation of the chain.
In addition, we neglect the possibility of self-intersections occurring in the configuration
after a pivot move is completed, since such configurations 
are negligible in the space ${\mathcal C}$.

With this algorithm, the topology of a polygon may change by a pivot move.
We estimate the knot type of polygons by calculating several topological invariants~\cite{DEGUCHI-TSURUSAKI.1993.01}, 
the value of the 
Alexander polynomial $\Delta_K(t)$ at $t = -1$ ~\cite{VOLOGODSKII-ET-AL.1974.01},
and the Vassiliev invariants of the second and third order, 
$v_2(K)$ and $v_3(K)$, respectively~\cite{POLYAK-VIRO.1994.01}.
The chance of miss identification is negligible for the 
simple knots we are concerned with.

We choose a certain conformation of a polygon as an initial state of the
Markov process. 
The mean deflection angle between adjacent segments is 
equal to $90^{\circ}$ in the thermal equilibrium.
We prepare seeds on the cubic lattice as an equilibrium conformations
of a polygon, exploiting the method used for the SAP on 
the cubic lattice~\cite{YAO-MATSUDA-TSUKAHARA-SHIMAMURA-DEGUCHI.2001.01}. 
Starting from a seed, we generate a sequence of polygons 
by applying the pivot moves repeatedly.
After discarding the initial $2000$ transient conformations,
we take samples of polygons at every $200$ pivot moves.

The simulation has been performed for polygons with the length $N$
between $100$ and $2200$ and for each of three knot 
types $\emptyset$, $3_1$ and $3_1\sharp4_1$.
We have collected $10^5$ polygons for each given length $N$ and for each given knot type and 
we have evaluated the radii of gyration (\ref{eq:all}) and (\ref{eq:k}).
We put a lower bound $N_{min}$ in fitting the simulation results 
using the scaling formula (\ref{eq:par}). 
Thus we make the fit in the range $N_{min} \leq N \leq N_{max}(=2200)$,
varying the value of $N_{min}$ from $100$ to $600$.
%

%
\section{Results of The Simulation}
\begin{figure}[htbp]
  \begin{center}
    \includegraphics[width=100mm]{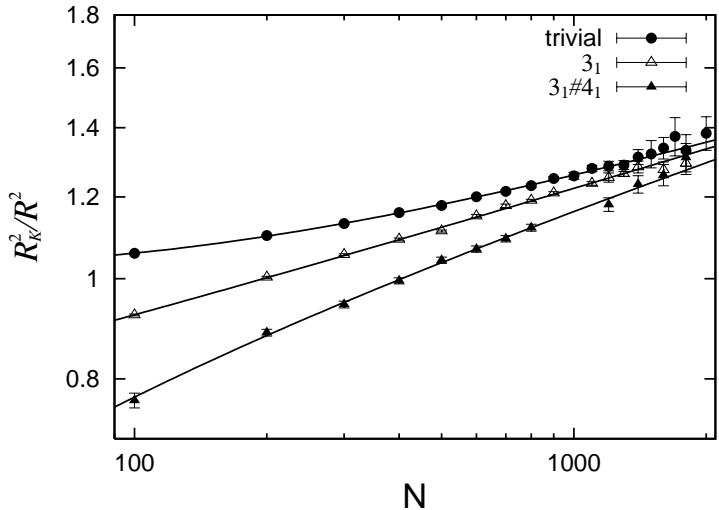}
    \caption{
       Simulation data on the ratios $R^2_{K}/R^2$ for the knot types 
       $\emptyset$ , $3_1$, $3_1\sharp4_1$
       and the best fit curves (solution 1) using (\ref{eq:par}).
       The scales are double-logarithmic.
    \label{fig:fitting1}}
  \end{center}
\end{figure}

We now discuss 
the results of our simulation for the three knot types.
The best fit curve for each knot type, together with the data on the ratio $R^2_{K}/R^2$, 
are shown in Fig.\ref{fig:fitting1}.
We have found that the scaling relation (\ref{eq:par}) 
fits the simulation data very well for all three cases 
in the range of $N$ from $N_{min} \gtrsim 100$ to $N_{max} = 2200$.
The estimated values of the parameters in (\ref{eq:par}) 
are shown in Table \ref{table:detailsfit1}.
The $\chi^2$ values remain small, 
$\chi^2 \simeq 1$ -- $2$ per datum, 
even if $N_{min}$ is lowered to $100$.
In the case of $3_1$, however, the $\chi^2$ values becomes small for
$N \gtrsim 600$.

The ratios $R^2_{K}/R^2$ increase as functions of $N$,
and they become larger than one for large $N$. 
The same behavior has been observed in other models, 
such as the Gaussian random polygon model~\cite{SHIMAMURA-DEGUCHI.2001.03} 
and the cylindrical SAP model~\cite{SHIMAMURA-DEGUCHI.2002.01}.
The above observation tells that 
the average size of polygons grows due to the topological constraint.
The exponent $\nu_{K}$ obtained in the fit are 
insensitive to the change of $N_{min}$, 
showing the validity of the scaling relation $R^2_{K} \thickapprox N^{2\nu_K}$.
The resulting values $2 \nu_K \simeq 1.11$ -- $1.16$ 
are consistent with $\nu_{SAW}$, 
supporting the assertion in the papers~\cite{DES_CLOIZEAUX.1981.01,DEUTSCH.1999.01}.

\begin{figure}[htbp]
  \begin{center}
    \includegraphics[width=100mm]{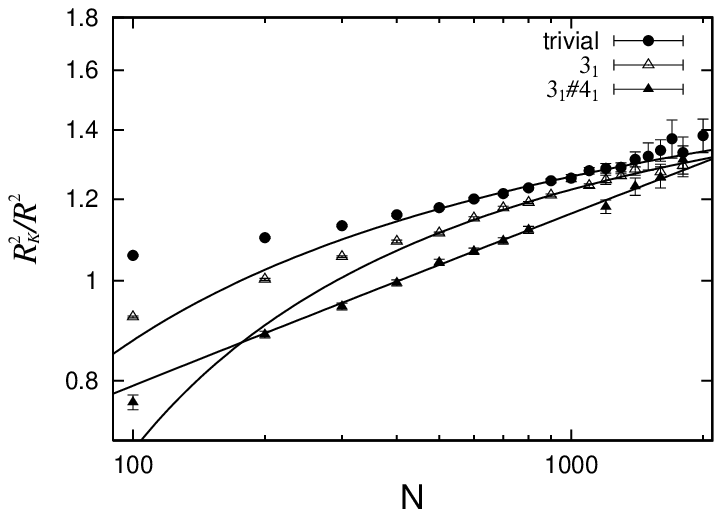}
    \caption{
       Simulation data on the ratio $R^2_{K}/R^2$ for the knot types 
       $\emptyset$ , $3_1$, $3_1\sharp4_1$
       and the curves of the second optimal fit (solution 2) using (\ref{eq:par}).
       The curves represent the second fit in which only data for $N \ge 600$ are used.
    \label{fig:fitting2}}
  \end{center}
\end{figure}

We have found another optimal fit to our simulation data, 
as shown in Fig.\ref{fig:fitting2}.
The curves represent the second fit 
in which only data for $N \gtrsim 600$ are used in Fig.\ref{fig:fitting2}.
The estimated values of the parameters are given in Table \ref{table:detailsfit2}.
The $\chi^2$ values for these fits grow rapidly 
as $N_{min}$ is lowered less than about $400$.
It implies that the formula (\ref{eq:par}) does not give a good approximation
in the range $N \lesssim 400$.

%
%
{\small
\begin{table}[htpb]
\begin{center}
\caption{
      The value of the parameters of the scaling relation
      (\ref{eq:par}) of the ratio $R^2_{K}/R^2$
      obtained from the best fit.}
\label{table:detailsfit1}
$trivial/all$\\
\begin{tabular}{lllll}
\hline
\hline
$N_{min}$  &  $A_{\emptyset}/A$  &  $\Delta B_{\emptyset}$  &  $2\Delta\nu_{\emptyset}$	 &  $\chi^2$ \\
\hline
100  &  0.471$\pm$0.014& 2.197$\pm$0.159& 0.133$\pm$0.004& 0.86\\
200  &  0.405$\pm$0.027& 3.172$\pm$0.429& 0.151$\pm$0.008& 0.69\\
400  &   0.376$\pm$0.079 & 3.779$\pm$1.619 & 0.159$\pm$0.024 & 0.70\\
600   &  0.242$\pm$0.154   & 7.812$\pm$6.457 & 0.207$\pm$0.068 & 0.679\\
\hline
\hline
\end{tabular}

$3_1/all$\\
\begin{tabular}{lllll}
\hline\hline
$N_{min}$  &  $A_{3_1}/A$  &  $\Delta B_{3_1}$  &  $2\Delta\nu_{3_1}$  &  $\chi^2$\\
\hline
100 &  0.494$\pm$0.030  & 0.258$\pm$0.269 & 0.130$\pm$0.008 & 6.26\\
200  & 0.455$\pm$0.064  & 0.722$\pm$0.782 & 0.140$\pm$0.017 & 6.56\\
400  & 0.624$\pm$0.221   &-1.281$\pm$1.973  &  0.104$\pm$0.0420 & 6.54\\
500 & 0.578$\pm$0.262&   -1.041$\pm$2.683&   0.114$\pm$0.053&  5.15\\
600  & 0.793$\pm$0.277   & -2.417$\pm$2.022   & 0.074$\pm$0.041 & 1.30 \\
\hline
\hline
\end{tabular}

$3_1 \sharp 4_1/all$\\
\begin{tabular}{lllll}
\hline\hline
$N_{min}$  &  $A_{3_1 \sharp 4_1}/A$  &  $\Delta B_{3_1 \sharp 4_1}$  &  $2\Delta\nu_{3_1 \sharp 4_1}$  &  $\chi^2$  \\
\hline
100&   0.460$\pm$0.048& -1.229$\pm$0.443& 0.140$\pm$0.013& 1.68\\
200 &  0.389$\pm$0.064& -0.369$\pm$0.831 &0.160$\pm$0.020& 1.52\\ 
400 &  0.378$\pm$0.159& -0.110$\pm$2.559 &0.163$\pm$0.049& 1.64  \\
600 &  0.293$\pm$0.335& 1.670$\pm$8.634& 0.192$\pm$0.128& 1.46\\
\hline
\hline
\end{tabular}
\end{center}
\end{table}
}

The second solution implies that the asymptotic behavior of the average size of polygons 
is scarcely affected by the topological constraint.
It gives $2 \nu_K \simeq 1.01$ -- $1.07$, 
which is not much different from $2 \nu = 1$.
Another noticeable difference is in the value of $A_K/A$.
$A_K/A < 1$ for the solution 1, whereas $A_K/A \simeq 1$ for the solution 2.

If we assume $A_{K}/A = 1$, then the effect of topological constraint could 
only appear in the coefficients $B_K$.
It is noted that the sign of $\Delta B_K$ is negative for each knot type with large statistical errors.
This observation supports the result derived by the perturbation argument in the limit of
large characteristic length~\cite{DEGUCHI-SHIMAMURA.2002.01}.

{\small
\begin{table}[htpb]
\caption{
      The value of the parameters of the scaling relation
      (\ref{eq:par}) of the ratio $R^2_{K}/R^2$
      obtained from the best fit.}
\label{table:detailsfit2}
\begin{center}
$trivial/all$\\
\begin{tabular}{lllll}

\hline\hline
$N_{min}$  &  $A_{\emptyset}/A$  &  $\Delta B_{\emptyset}$  &  $2\Delta\nu_{\emptyset}$  &  $\chi^2$\\
\hline
200&1.720$\pm$0.464&-3.731$\pm$0.915&-0.029$\pm$0.035&49.42\\
400&1.736$\pm$0.585&-4.820$\pm$1.37&-0.023$\pm$0.042&7.24\\
500&1.359$\pm$0.270&-5.345$\pm$0.844&0.012$\pm$0.024&1.91\\
600&1.203$\pm$0.514&-3.884$\pm$2.214&0.026$\pm$0.050&1.66\\
\hline
\hline
\end{tabular}

$3_1/all$\\
\begin{tabular}{lllll}
\hline\hline
$N_{min}$  &  $A_{3_1}/A$  &  $\Delta B_{3_1}$  &  $2\Delta\nu_{3_1}$  &  $\chi^2$\\
\hline
200 &  1.273$\pm$0.340 & -3.867$\pm$0.941 & 0.012$\pm$0.034 & 61.19 \\
400 &  1.117$\pm$0.370 & -4.181$\pm$1.461 & 0.034$\pm$0.040 & 9.17\\
500  & 1.286$\pm$0.261&  -5.107$\pm$0.882 & 0.019$\pm$0.025 & 1.92\\ 
600  & 1.095$\pm$0.360 & -4.180$\pm$1.684 & 0.037$\pm$0.039 & 1.50  \\
\hline
\hline
\end{tabular}

$3_1 \sharp 4_1/all$\\
\begin{tabular}{lllll}
\hline\hline
$N_{min}$  &  $A_{3_1 \sharp 4_1}/A$  &  $\Delta B_{3_1 \sharp 4_1}$  &  $2\Delta\nu_{3_1 \sharp 4_1}$  &  $\chi^2$  \\
\hline
200 &  0.926$\pm$0.306& -4.056$\pm$1.137& 0.050$\pm$0.043& 12.22 \\
400 &  0.997$\pm$0.432& -4.992$\pm$1.787& 0.047$\pm$0.053& 3.80\\
600 &  0.877$\pm$0.827& -5.021$\pm$4.623& 0.066$\pm$0.111  & 2.67\\
\hline
\hline
\end{tabular}
\end{center}
\end{table}
}


To summarize, we have confirmed that the simulation data on the average size $R_K^2$ 
of ring polymers can be fitted by the scaling relation (\ref{eq:par}).
We have found two solutions.
The best fit solution (solution 1) is good for $100 \le N \le 2200$ with quite small $\chi^2$ 
($\chi^2 \simeq 1$) while the second solution (solution 2) becomes good for $N_{min} \gtrsim 600$.
The two solutions lead to two different interpretations regarding 
the effect of fixed knot topology on the average size of ring polymers as their length $N$ increases.

\section*{Acknowledgment}
The authors are grateful to M. K. Shimamura for helpful discussions.
A.Y. and H.M. are supported by Research Assistant Fellowship of Chuo University.
T.I. are supported  partially by Research grant of Japanese Ministry of Education and Science, 
Kiban C (2), 
and Chuo University grant for special research.


\begin{thebibliography}{10}

\bibitem{DOI-EDWARDS.1986.01}
M.~Doi and S.~F. Edwards.
\newblock {\em The Theory of Polymer Dynamics}.
\newblock Oxford Sci. Pub., New York, 1986.

\bibitem{DELBRUCK.1962.01}
M.~Delbr{\"{u}}ck.
\newblock Knotting problems in biology.
\newblock {\em Proc. Symp. Appl. Math.}, 4:55--63, 1962.

\bibitem{SUMNERS.1995.01}
D.~W. Sumners.
\newblock Lifting the curtain: using tolopgy to probe the hidden action of
  emzymes.
\newblock {\em Notices of the AMS}, 42:528--537, 1995.

\bibitem{STASIAK-ET-AL.1996.01}
A.~Stasiak, V.~Katritch, J.~Bednar, D.~Michoud, and J.~Dubochet.
\newblock Electrophoretic mobility of dna knots.
\newblock {\em Nature}, 384:122--122, 1996.

\bibitem{DES_CLOIZEAUX-METHA.1979.01}
J.~des Cloizeaux and M.~L. Metha.
\newblock Topological constraints on polymer rings and critical indices.
\newblock {\em J. Physique}, 40:665--670, 1979.

\bibitem{QUAKE.1994.01}
S.~R. Quake.
\newblock Topological effects of knots in polymers.
\newblock {\em Phys. Rev. Lett.}, 73:3317--3320, 1994.

\bibitem{JANS_VAN_RENSBURG-WHITTINGTON.1991.01}
E.~J.~Janse van Rensburg and S.~G. Whittington.
\newblock The dimensions of knotted polygons.
\newblock {\em J. Phys. A}, 24:3935--3948, 1991.

\bibitem{ORANDINI-TESI-JANSE_VAN_RENSBURG-WHITTINGTON.1998.01}
E.~Orlandini, M.~Tesi, E.~J.~Janse van Rensburg, and S.~G. Whittington.
\newblock Asymptotics of knotted lattice polygons.
\newblock {\em J. Phys. A}, 31:5953--5967, 1998.

\bibitem{DEUTSCH.1999.01}
J.~M. Deutsch.
\newblock Equilibrium size of large ring molecules.
\newblock {\em Phys. Rev. E}, 59:R2539--41, 1999.

\bibitem{SHIMAMURA-DEGUCHI.2001.02}
M.~K. Shimamura and T.~Deguchi.
\newblock Gyration radius of a circular polymer under a topological constraint
  with excluded volume.
\newblock {\em Phys. Rev. E}, 64:R020801, 2001.

\bibitem{SHIMAMURA-DEGUCHI.2001.03}
M.~K. Shimamura and T.~Deguchi.
\newblock Anomalous finite-size effects for the mean-squared gyration radius of
  gaussian random knots.
\newblock {\em J. Phys. A}, 35:L241--L246, 2002.

\bibitem{SHIMAMURA-DEGUCHI.2002.01}
M.~K. Shimamura and T.~Deguchi.
\newblock Finite-size and asymptotic behaviors of the gyration radius of
  knotted cylindrical self-avoiding polygons.
\newblock {\em Phys. Rev. E}, 65:051802, 2002.

\bibitem{DE_GENNES.1972.01}
P.~G. de~Gennes.
\newblock Exponents for the excluded volume problem as derived by the wilson
  method.
\newblock {\em Phys. Lett. A}, 38:339--340, 1972.

\bibitem{WILSON.1974.01}
K.~G. Wilson and J.~B. Kogut.
\newblock The renomalization group and the $\varepsilon$ expansion.
\newblock {\em Phys. Rep. C}, 12:75--199, 1974.

\bibitem{DES_CLOIZEAUX.1974.01}
J.~des Cloizeaux.
\newblock Lagrangian theory for a self-avoiding random chain.
\newblock {\em Phys. Rev. A}, 10:1665--1669, 1974.

\bibitem{LIPKIN.1981.01}
M.~Lipkin, Y.~Oono, and K.~F. Freed.
\newblock Confrmation space renormalization of polymers. 4. equilibrium
  properties of the simple ring polymer using gell-mann-low type
  renormalization group theory.
\newblock {\em Macromolecules}, 14:1270--1277, 1981.

\bibitem{PRENTIS.1981.01}
J.~J. Prentis.
\newblock Spatial correlations in a self-repelling ring polymer.
\newblock {\em J. Chem. Phys.}, 76:1574--1583, 1981.

\bibitem{WHITTINGTON.1992.01}
S.~G. Whittington.
\newblock Topology of polymers.
\newblock {\em Proc. Symp. Appl. Math.}, 45:73--95, 1992.

\bibitem{DES_CLOIZEAUX.1981.01}
J.~des Cloizeaux.
\newblock Ring polymers in soluion: topological effects.
\newblock {\em J. Physique Lett.}, 42:L433--L436, 1981.

\bibitem{GROSBERG.2000.01}
A.~Yu. Grosberg.
\newblock Critical exponents for random knots.
\newblock {\em Phys. Rev. Lett.}, 85:3858, 2000.

\bibitem{MADRAS-SOKAL.1987.01}
N.~Madras and A.~D. Sokal.
\newblock The pivot algorithm: a highly efficient monte carlo method for the
  self-avoiding walk, 1987.

\bibitem{MADRAS-ORLITSKY-SHEPP.1990.01}
N.~Madras, A.~Orlitsky, and L.~A. Shepp.
\newblock Monte carlo generation of self-avoiding walks with fixed endpoints
  and fixed length.
\newblock {\em J. Stat. Phys.}, 58:159--183, 1990.

\bibitem{DEGUCHI-TSURUSAKI.1993.01}
T.~Deguchi and K.~Tsurusaki.
\newblock A new algorithm for numerical calculation of link invariants.
\newblock {\em Phys. Lett. A}, 174:29--37, 1993.

\bibitem{VOLOGODSKII-ET-AL.1974.01}
A.~V. Vologodskii et~al.
\newblock The knot problem in statistical mechanics of polymer chains.
\newblock {\em Sov. Phys. JETP}, 39:1059--1063, 1974.

\bibitem{POLYAK-VIRO.1994.01}
M.~Polyak and O.~Viro.
\newblock Gauss diagram formulas for vassiliev invariants.
\newblock {\em Int. Math. Res. Not.}, (11):445--453, 1994.

\bibitem{YAO-MATSUDA-TSUKAHARA-SHIMAMURA-DEGUCHI.2001.01}
A.~Yao, H.~Matsuda, H.~Tsukahara, M.~K. Shimamura, and T.~Deguchi.
\newblock On the dominance of trivial knots among saps on a cubic lattice.
\newblock {\em J. Phys. A}, 34:7563--7577, 2001.

\bibitem{DEGUCHI-SHIMAMURA.2002.01}
T.~Deguchi and M.~K. Shimamura.
\newblock Topological effects on the average size of random knots.
\newblock {\em Contemporary Mathematics}, 304:93--114, 2002.

\end{thebibliography}

\end{document}